\documentclass[12pt]{article}

\usepackage{epsfig}
\begin{document}

\vspace*{2.0cm}

\begin{center}
{\large {\bf Astrophysical reaction rate for
$\alpha(\alpha n,\gamma)$$^{9}$Be\\
by photodisintegration\\}  }
\vspace*{2.0cm}
K. Sumiyoshi$^{a,}$\footnote{e-mail: sumi@numazu-ct.ac.jp},
H. Utsunomiya$^{b,}$\footnote{e-mail: hiro@konan-u.ac.jp},
S. Goko$^{b,}$\footnote{e-mail: gokou@konan-u.ac.jp},
and T. Kajino$^{c,}$\footnote{e-mail: kajino@nao.ac.jp} \\
\vspace*{0.5cm}

$^{a}$Numazu College of Technology, \\
Ooka 3600, Numazu, Shizuoka 410-8501, Japan \\
$^{b}$Department of Physics, Konan University, \\
Okamoto 8-9-1, Higashinada-ku, Kobe, Hyogo 658-8501, Japan \\
$^{c}$National Astronomical Observatory, \\
Osawa 2-21-1, Mitaka, Tokyo 181-8588, Japan \\
\end{center}

\vspace*{0.5cm}
\newpage

\begin{abstract}
We study the astrophysical reaction rate for the formation of
$^{9}$Be through the three body reaction $\alpha(\alpha n,\gamma)$.
This reaction is one of the key reactions which could bridge the mass gap 
at A = 8 nuclear systems to produce intermediate-to-heavy mass elements
in alpha- and neutron-rich environments such as r-process nucleosynthesis
in supernova explosions, s-process nucleosynthesis in asymptotic giant 
branch (AGB) stars, and primordial nucleosynthesis in baryon inhomogeneous
cosmological models.
To calculate the thermonuclear reaction rate in a wide range of 
temperatures,
we numerically integrate the thermal average of cross
sections assuming a two-steps formation through a metastable
$^{8}$Be, $\alpha + \alpha$ 
$\rightleftharpoons$ $^{8}$Be(n,$\gamma$)$^{9}$Be.
Off-resonant and on-resonant contributions 
from the ground state 
in $^{8}$Be are taken into account.
As input cross section, we adopt the latest
experimental data by photodisintegration of $^{9}$Be
with laser-electron photon beams, which 
covers all relevant resonances in $^{9}$Be.  
Experimental data near the neutron threshold are added 
with $\gamma$-ray flux corrections 
and a new least-squares analysis is made to deduce 
resonance parameters in the Breit-Wigner formulation.  
Based on the photodisintegration cross section, 
we provide the reaction rate for 
$\alpha(\alpha n,\gamma)^{9}$Be 
in the temperature range 
from T$_{9}$=10$^{-3}$ to T$_{9}$=10$^{1}$ 
(T$_{9}$ is the temperature in units of $10^{9}$ K) 
both in the tabular
form and in the analytical form for potential usage in 
nuclear reaction network calculations.
The calculated reaction rate is compared with the reaction 
rates of the CF88 and the NACRE compilations.  
The CF88 rate, which is based on the photoneutron cross section 
for the 1/2$^{+}$ state in $^{9}$Be by Berman {\it et al.}, is valid 
at $T_{9} > 0.028$ due to lack of the off-resonant contribution.  
The CF88 rate differs from the present rate by a factor of two 
in a temperature range $T_{9} \geq 0.1$.  
The NACRE rate, which adopted different sources of experimental information 
on resonance states in $^{9}$Be, is 4--12 times larger than 
the present rate at $T_{9} \leq 0.028$, 
but is consistent with the present rate to within $\pm 20 \%$ 
at $T_{9} \geq 0.1$.  
\end{abstract}

\newpage

\section{Introduction}

The interplay between nuclear physics and astrophysics is
essential to clarify the origin of elements in the Universe.
Scenarios of the nucleosynthesis in stars rely on
the determination of important nuclear reactions
and structures along the reaction path \cite{Bur57}.
A long way to create heavy elements 
starting from light elements (p, n, $\alpha$) 
has to pass
through the gaps in mass number at A=5 and A=8.  
The bridge to heavy elements such as $^{12}$C invokes
slow three-body reactions and then links to heavier 
elements up to iron elements and beyond.
The triple alpha reaction for the formation of $^{12}$C
is usually the key reaction in stars at the helium burning stage.

In a neutron-rich environment, however, there is an 
important three-body reaction.
If neutrons are abundant enough together with alphas,
the formation of $^{9}$Be can proceed through
the reaction $\alpha(\alpha n,\gamma)$.
This reaction followed by $^{9}$Be($\alpha$,n)$^{12}$C
can compete with the triple alpha reaction
and may dominate over, depending on the astrophysical environment.

The importance of the $\alpha \alpha n$ reaction has been
pointed out in the r-process in the hot bubble of a
core-collapse supernova \cite{Mey92,Woo94}.
After the core bounce in the supernova, a nascent neutron
star is formed at the center.
The outer material of this proto-neutron star is heated by intense neutrinos
emitted from the proto-neutron star's inside and is ejected
as a wind.
A high-entropy bubble in this neutrino-driven wind
is thought to be an ideal site for the r-process nucleosynthesis.
Because of the high temperature and the low density, the matter
is dissociated into neutrons and protons once, and
they reassemble into light elements, mostly alphas.
The nucleosynthesis starts from alphas and remaining
neutrons when the temperature decreases below T$_{9} \sim 5$
in the expanding matter.
Since the triple alpha reaction is too slow 
at the temperature-density conditions in the hot bubble, 
the $\alpha \alpha n$
reaction forms a bypass flow
connecting to $^{9}$Be($\alpha$,n)$^{12}$C.
Therefore, the determination of the reaction rate
for the $^{9}$Be formation is essential for the initial 
condition of the r-process, namely, the amounts of 
seed elements and remaining neutrons.  
Because of the high entropy (i.e. low density), 
the three body reaction proceeds slowly, resulting in 
seed elements with a high neutron-to-seed ratio 
for a successful r-process up to A=200.  

In recent studies of the r-process in the neutrino-driven
wind, the importance of this reaction is more evident.
It was shown that the r-process can be successful
in the wind flow with a very short expansion time
and a moderate entropy \cite{Ots00, Sum00}.
In this rapid expansion, there is essentially no time
for the triple alpha reaction to set in and, instead, 
the $\alpha \alpha n$ reaction becomes again crucial
for the r-process.
The necessary expansion time is determined in 
comparison with the reaction time scale of
the $\alpha \alpha n$ reaction \cite{Ots00}.
A more recent study reveals that the reaction flow
can go through the neutron-rich region of 
light elements in the rapidly expanding
wind \cite{Ter01, Ter01a}.
Therefore, the competition of the $\alpha \alpha n$
reaction with other bypass reactions 
in the short expansion time should be carefully considered.  

The $\alpha \alpha n$ reaction may play an important
role in other astrophysical environments.  
In the systematic study of alpha-rich freeze-out, 
the $\alpha \alpha n$ reaction may regulate the
production of elements up to A=100 even if it does
not lead to the r-process \cite{Woo92, How93}.
In the neutron-rich He layer of low-mass 
asymptotic giant branch (AGB) stars, 
the $\alpha \alpha n$ reaction might be able to
take place below T$_{9}$=0.3.
This may change the amount of neutrons to determine
the abundance of s-process elements \cite{Iwa01}.
In the inhomogeneous Big-Bang nucleosynthesis,
the $\alpha \alpha n$ reaction might contribute
to the production of heavier elements beyond A=8
in neutron-rich zones
at temperature around T$_{9}$=0.1 \cite{Ori97}.

Most of nucleosynthesis studies treating the
$\alpha \alpha n$ reaction adopt the astrophysical
reaction rate from the compilation of CF88 \cite{Cau88}.  
A more recent evaluation of the $^{9}$Be formation
rate can be found in the NACRE compilation of the
thermonuclear reaction rates \cite{Ang99}.

Since the direct measurement of the 
$\alpha(\alpha n,\gamma)$$^{9}$Be reaction 
is impossible, the experimental
efforts to provide information on
excited states in $^{9}$Be have been made
(see \cite{Ang99, Uts01} and the references therein).
A ternary process hardly plays a role
in the formation of $^{9}$Be.  
Instead, the reaction reflects nuclear structure 
of $^{9}$Be with large neutron-decay widths of 
low-lying states in $^{9}$Be.  
The reaction is described by $\alpha + \alpha$ 
$\rightleftharpoons$ $^{8}$Be(n,$\gamma$)$^{9}$Be, 
which takes place during the lifetime of $^{8}$Be 
($\sim$ 10$^{-16}$ s) in such suitable astrophysical 
site as the neutrino-driven wind.  
Therefore, the resonance structure of $^{9}$Be
near the threshold for $^{8}$Be+n at 1.665 MeV 
is of great importance \cite{Ajz88}.
It is noted that the reaction $\alpha + n$ 
$\rightleftharpoons$ $^{5}$He($\alpha$,$\gamma$)$^{9}$Be
with the lifetime of $^{5}$He ($\sim$ 10$^{-21}$ s) 
can make an additional contribution to the formation 
of $^{9}$Be \cite{Buc97, Buc01}.  

Recently, a systematic measurement of the 
$^{9}$Be($\gamma$,n)$^{8}$Be reaction was made 
with laser-electron photon beams 
in the energy range of astrophysical relevance \cite{Uts01}.
This study has provided 
photoneutron cross sections as a function
of $\gamma$-ray energy.  
The data enables us
to derive the neutron capture cross section of $^{8}$Be 
with the detailed-balance theorem, 
which is necessary in the calculation of the reaction rate
for the $^{9}$Be formation.
This systematic treatment of all the resonances
is preferable in contrast with the preceding treatment 
which adopted 
different sources of experimental information on the resonances 
together with some assumptions.  

The brief report \cite{Uts01} provided the reaction
rate of $< \alpha \alpha n >$
as a function of temperature.
However, the reaction rate was not tabulated,
but shown only graphically as a ratio
with respect to the NACRE compilation.
The temperature range was limited between T$_{9}$=0.1
and T$_{9}$=10, where only the 
on-resonant contribution from $^{8}$Be was 
taken into account.  
In contrast, the NACRE compilation treated a broader 
temperature range down to T$_{9}$=10$^{-3}$ for
possible astrophysical applications.
At temperatures below T$_{9}$=0.028, 
the effective energy window for the $\alpha \alpha n$ reaction 
is far below the resonance energy of
$^{8}$Be.
In this case, the off-resonant contribution 
for the formation of 
$^{8}$Be from $\alpha + \alpha$ becomes important.  
This contribution was first shown to be important
in the determination of the triple alpha reaction in
accreting white dwarfs and neutron stars \cite{Nom85}.

In this paper, we calculate the reaction
rate for the $^{9}$Be formation taking into account
both on- and off-resonant contributions 
based on the latest experimental information 
on photoneutron cross sections for $^{9}$Be.  
It is to be noted that we add data points 
immediately above the neutron threshold 
since the brief report \cite{Uts01}.  
We provide the 
reaction rate in the wide range of temperatures
from T$_{9}$=10$^{-3}$ to T$_{9}$=10$^{1}$
in the analytical fitting formula
as well as in the tabular form.

We arrange this article as follows.
In section \ref{sec:review},
we briefly review the preceding efforts
to derive the $\alpha \alpha n$ reaction rate
from the point of view of experimental
information and experimental methods adopted.  
In section \ref{sec:experiment},
we describe the photoneutron
cross section for $^{9}$Be adopted to derive
the reaction rate for the neutron capture of $^{8}$Be.
The description of the experimental method 
is limited to minimum unless it is associated with 
added data and a new analysis.  
In section \ref{sec:theory},
we explain the theoretical
method to derive the $\alpha \alpha n$
reaction rate from the experimental data.
In section \ref{sec:result},
we provide the numerical results 
of the reaction rate in a tabular form 
and plots.
We also provide the fitting formulae of 
the reaction rate.  
We compare the result with the reaction rate 
in the CF88 and NACRE compilations.  
The article is closed with a summary 
in section \ref{sec:summary}.

\section{Resonances of $^{9}$Be}\label{sec:review}

Fowler, Caughlan, and Zimmerman \cite{FCZII} presented the reaction 
rate of $\alpha$($\alpha$n, $\gamma$)$^{9}$Be in their second volume 
of {\it Thermonuclear Reaction Rates} published in 1975 (hereafter 
referred to as FCZ II).  
The rate has not been revised in the third volume (1983) \cite{HFCZ} 
of the series and subsequently passed on to the two compilations of 
Caughlan, Fowler, Harris, and Zimmerman (1985) \cite{CFHZ85} and of 
Caughlan and Fowler (1988) \cite{Cau88} (hereafter referred to as 
CF88) without modification.  It was nearly a quarter century after 
the FCZ II that Angulo {\it et al.} \cite{Ang99} updated the reaction 
rate (hereafter referred to as NACRE).  

Early measurements of photoneutron cross sections for $^{9}$Be were 
carried out with monochromatic $\gamma$ rays from radioactive 
isotopes \cite{Rus48, Ham53, Edg57, Gib59,Joh67} and with 
bremsstrahlung \cite{Jak61,Ber67}.  The data of Berman {\it et al.} 
exhibit the maximum cross section 1.6 mb at 6 keV above the n + 
$^{8}$Be threshold (E$_{t}$ = 1.665 MeV) and decrease to 1.2 mb at 40 
keV \cite{Ber67}.  Parametrizing the photoneutron cross sections by 
$\sigma_{\gamma n}$ = 1.6 exp(-4E) mb (E in MeV) with E = 
E$_{\gamma}$ - E$_{t}$ \cite{Ang99} resulted in the CF88 reaction 
rate N$_{A}^{2}$$<\alpha \alpha n>$ as

\begin{equation}
{N}_{A}^{2} \left\langle{\alpha \alpha \mit n}\right\rangle = 
2.59 \times {10}^{- 6} {\exp ( - 1.062 / {T}_{9}) 
\over [{T}_{9}^{2} (1 + 0.344 {T}_{9})]},
\label{rate1}
\end{equation}
where T$_{9}$ gives temperature as T = T$_{9} \times$ 10$^{9}$ K and 
N$_{A}$ is Avogadro's number.  
The low energy experimental data were recently investigated 
in a semi-classical model to derive the $\alpha$($\alpha$n, $\gamma$)$^{9}$Be 
reaction rate \cite{Efr98}.

Woosley and Hoffman discussed the reaction rate in a narrow level 
approximation for two lowest states (the 1/2$^{+}$ and 5/2$^{-}$ states) 
in the n + $^{8}$Be channel.  It was pointed out that when the resonance parameters 
from the (e,e$'$) data of Clerc {\it et al.} \cite{Cle68} are employed (i.e., E$_{x1}$ 
= 1.78 $\pm$ 0.03 MeV and $\Gamma_{\gamma 1}$ = 0.30 $\pm$ 0.12 eV 
for 1/2$^{+}$ and E$_{x2}$ = 2.44 $\pm$ 0.02 MeV and $\Gamma_{\gamma 
2}$ = 0.089 $\pm$ 0.01 eV for 5/2$^{-}$), the rate is consistent with 
the CF88 rate to within 30 \% in the range T$_{9}$ = 1 - 3 
\cite{Woo92}.  The narrow level approximation holds for the 5/2$^{-}$ 
state with $\Gamma$ = 0.77 $\pm$ 0.15 keV, but not for the 1/2$^{+}$ 
state with $\Gamma$ = 217 $\pm$ 10 keV \cite{Ajz88}.  In the 
discussion, the discrepancy in E$_{x1}$ was also pointed out between Ref.~\cite{Cle68} and 
the latest compilation of nuclear physics data \cite{Ajz88} which adopted 
E$_{x1}$ = 1.684 MeV $\pm$ 7 keV from a different (e,e$'$) data of 
Kuechler {\it et al.} \cite{Kue87}.  

After the FCZ II, some new measurements of $\sigma_{\gamma n}$ were 
carried out with both monochromatic $\gamma$ rays from radioactive 
isotopes \cite{Fuj82} and with bremsstrahlung  \cite{Hug75,Gor92}.  An 
R-matrix fit to the $\sigma_{\gamma n}$ of Fujishiro {\it et al.} 
\cite{Fuj82} in one level approximation gave parameters 
B = 1.06$^{+0.19}_{-0.16}$ mb, $\epsilon_{R}$ = 192 keV, and E$_{R}$ = 67 keV 
\cite{Bar83} which can be converted to B(E1\,$\downarrow$) = 
0.106$^{+0.019}_{-0.016}$ e$^{2}$fm$^{2}$, E$_{x}$ = 1.733 MeV, and $\Gamma$ 
$\cong$ $\Gamma_{n}$ = 227 keV, respectively.  The last quantity was 
evaluated from ${\Gamma }_{n} = 2 \sqrt {{\epsilon }_{R }({E}_{\gamma 
} - {E}_{t})}$ with ${E}_{\gamma } - {E}_{t}$ = 67 keV.   A major 
difference between the photonuclear reaction and the electron scattering 
is that the former B(E1\,$\downarrow$) \cite{Fuj82}
is a factor of two lager than the latter one (B(E1\,$\downarrow$) = 
0.050 $\pm$ 0.020 \cite{Cle68} and B(E1\,$\downarrow$) = 0.054 $\pm$ 
0.004 \cite{Kue87}).  According to the Barker's analysis, the main 
reason for the discrepancy may be due to a potential problem of 
background subtraction in the electron scattering.  Barker pointed out  
that the (e,e$'$) result may represent not a full but a partial strength 
of B(E1\,$\downarrow$). 

The NACRE rate of Angulo {\it et al.} is based on three states in 
$^{9}$Be, i.e., the 1/2$^{+}$, 1/2$^{-}$, and 5/2$^{+}$ states, 
neglecting the 5/2$^{-}$ state.  Resonance parameters (E$_{x}$, 
$\Gamma_{n}$, and $\Gamma_{\gamma}$) employed for the 1/2$^{+}$ state 
were taken from the $\sigma_{\gamma n}$ measurements of 
Fujishiro {\it et al.} \cite{Fuj82,Bar83}, where $\Gamma_{\gamma}$ was 
0.51 $\pm$ 0.10 eV.  
For the broad 1/2$^{-}$ state at E$_{x}$ = 2.78 $\pm$ 0.12 MeV with 
$\Gamma$ = 1080 $\pm$ 110 keV \cite{Ajz88},  $\Gamma_{\gamma}$ = 1 
W.u. (0.45 eV) for an M1 decay was assumed with a 80 \% uncertainty  
\cite{Des89}.  For the 5/2$^{+}$ state at E$_{x}$ = 3.049 MeV $\pm$ 9 
keV with $\Gamma$ = 282 $\pm$ 11 keV \cite{Ajz88}, $\Gamma_{\gamma}$ 
= 0.90 $\pm$ 0.45 eV which is much larger than the (e,e$'$) result (0.30 $\pm$ 0.25 eV) of 
Clerc {\it et al.} was adopted with the increased uncertainty based 
on the $\sigma_{\gamma n}$ data of Hughes {\it et al.} \cite{Hug75}.  

Recently, Utsunomiya {\it et al.} made a new measurement of 
$\sigma_{\gamma n}$ for $^{9}$Be in the range of E$_{\gamma}$ = 1.78 
- 6.11 with quasi-monochromatic $\gamma$ rays produced by means of 
inverse Compton scattering of laser photons \cite{Uts01}.  The data 
essentially include all states of astrophysical interest.  A 
least-squares fit to the data was carried out to extract the Breit-Wigner 
resonance parameters.  Using the ($\gamma$,n) cross sections, the 
$\alpha$($\alpha$n, $\gamma$)$^{9}$Be reaction rate was evaluated.  
This evaluation was based only on the resonant contribution to the 
reaction $\alpha$ + $\alpha$ $\rightleftharpoons$ $^{8}$Be with 
$\Gamma_{\alpha} = 6.8 \pm 1.7$ eV.  Consequently, the reaction rate 
is valid only at $T_{9} > 0.028$ 
as is the case for the triple alpha reaction \cite{Nom85}.

The 5/2$^{-}$ state was treated differently in the NACRE rate and the 
rate of Utsunomiya {\it et al.}  It was completely omitted in the 
former, whereas it was taken into account in the latter assuming its 
100 \% decay to the $^{8}$Be(0) + n channel.  The 5/2$^{-}$ state in 
$^{9}$Be was investigated in the $^{7}$Li($^{3}$He,p)$^{9}$Be 
reaction \cite{Chr66} and in $\beta$ decay of $^{9}$Li \cite{Che70} 
and found to decay 7.5 $\pm$ 1.5 \% and 6.4 $\pm$ 1.2 \% to 
$^{8}$Be(0) + n, respectively.  It is reported that the 5/2$^{-}$ 
state decays dominantly to $^{5}$He(0) + $\alpha$ \cite{Ren71}.  The 
5/2$^{-}$ state with $\Gamma$ = 0.77 $\pm$ 0.15 keV lies at E$_{x}$ = 
2.4294 MeV $\pm$ 1.3 keV below the $^{5}$He + $^{4}$He threshold energy 
(2.467 MeV).  However, since the ground state in $^{5}$He has a large 
width (0.60 $\pm$ 0.02 MeV) \cite{Ajz88}, the state can decay into 
the low-energy tail of the ground state.  

A potential contribution of 
the $^{5}$He(0) + $\alpha$ channel to the reaction rate was investigated 
by Buchmann \cite{Buc97} where a 5 \% branching of the 5/2$^{-}$ state 
to the n + $^{8}$Be decay channel was taken.  Summing up all possible contributions through 
the $^{5}$He + $\alpha$ channel not only from the 5/2$^{-}$ state but also 
from the 1/2$^{+}$ and 1/2$^{-}$ states resulted in a reaction rate that is
one order of magnitude smaller than through the n + $^{8}$Be channel at 
high temperatures, T$_{9}$ $\geq$ 2  (Fig. 9 of Ref.~\cite{Buc97}).  

In a recent experiment of $\beta$-delayed particle decay of $^{9}$C, 
a large background continuum in $^{9}$B which decays to the $^{5}$Li channel 
was observed above 2.9 MeV \cite{Get00}.  An R-matrix fit revealed 
that a low energy part of the continuum consists of a tail of 
the ground state \cite{Buc01}.  The charge-symmetry analogue of 
this background continuum in $^{9}$Be may contribute to 
the $\alpha(\alpha n, \gamma)^{9}$Be reaction rate.  
An attempt to estimate the contribution of the analogue background 
continuum to the $\alpha$($\alpha$n, $\gamma$)$^{9}$Be reaction rate 
through the $^{5}$He + $\alpha$ channel was made, showing that 
it possibly dominates the rate for high temperatures 
above T$_{9} \approx 4$ \cite{Buc01,Buc02}.  
This result may be qualitatively consistent with the result of 
a recent microscopic three-cluster study of photoneutron cross 
sections for $^{9}$Be which showed that the $^{5}$He + $\alpha$ channel 
is of growing importance above E$_{\gamma}$ $\approx$ 4 MeV \cite{Des01}.  

In the present paper, we evaluate the $\alpha$($\alpha$n, $\gamma$)$^{9}$Be 
reaction rate through the $^{8}$Be + n channel, including a fractional 
contribution of the 5/2$^{-}$ state. 
The contribution through the $^{5}$He + $\alpha$ channel is not included 
because of a lack of sufficient experimental information 
on the $^{5}$He + $\alpha$ channel.  
Clearly, the evaluation of the stellar reaction rate at the next level 
must await a new experimental outcome for the decay of relevant states 
in $^{9}$Be through the $^{5}$He + $\alpha$ channel 
(particle- and $\gamma$- widths and branching ratios of decays) 
which is expected in the proposed experiment of 
$\beta$-delayed particle decay of $^{9}$Li \cite{Buc97,Buc01}.  

\section{Photodisintegration of $^{9}$Be}\label{sec:experiment}

\subsection{Experiment}

We limit a description of the experiment to minimum 
and concentrate on new facets of the data and data analysis.  
Many of the details can be 
found in \cite{Uts01}.  The experiment was performed at the National 
Institute of Advanced Industrial Science and Technology (AIST) 
which is reorganized from 
the Electrotechnical Laboratory.  Quasi-monochromatic $\gamma$-ray beams were 
generated by means of inverse Compton scattering of YLF: Nd laser 
photons with $\lambda$ = 1053 nm (1.2 eV) from relativistic electrons 
stored in an accumulator ring TERAS.  The energy of the 
laser-electron photon beams (LEPBs) was varied in the energy range of 
1.69 - 6.11 MeV by changing the electron beam energy from 308 to 587 
MeV.  The LEPBs were collimated into a spot 2 mm in diameter with a 
20 cm thick Pb block at 550 cm from the head-on collision point.  The 
beam intensity which gradually decreased with the electron beam 
current was in the range of 10$^{5}$ - 10$^{4}$ cps.  A 4 cm thick 
$^{9}$Be was irradiated and neutrons were detected 
with four BF$_{3}$ counters embedded in a 30 cm polyethylene cube.  
A BGO detector was used as a flux monitor of the LEPB.  

Experimentally two reaction channels of n + $^{8}$Be and $^{5}$He + $\alpha$
cannot be distinguished.  
Different cross sections may result depending on the reaction channels 
because the neutron detection efficiency may be 
different in the two channels.  This complication 
could occur for the 5/2$^{-}$ state which decays to the two channels, whereas 
other states are free from such complication because they are known to 
predominantly decay to the n + $^{8}$Be channel \cite{Ajz88}.  
The average energy of neutrons emerging from the decay of the 5/2$^{-}$ state 
to the two channels was calculated.  The energy of this state is 
well defined at 2.4294 MeV ($\pm$ 1.3 keV) with $\Gamma$ = 0.77 keV ($\pm$ 0.15 keV).
When this state decays to the n + $^{8}$Be channel at 1.665 MeV, the average laboratory 
energy of neutrons is 382 keV.  On the other hand, this state can decay to the 
$^{5}$He + $\alpha$ channel at 2.467 MeV with the low-energy tail of $\Gamma$ ($^{5}$He) = 0.60 MeV ($\pm$ 0.02 MeV).  
The average energy of neutrons resulting from $^{5}$He was 
calculated to be 156 keV.  

\begin{figure}[t]
\begin{center}
\vspace*{5cm}
\epsfig{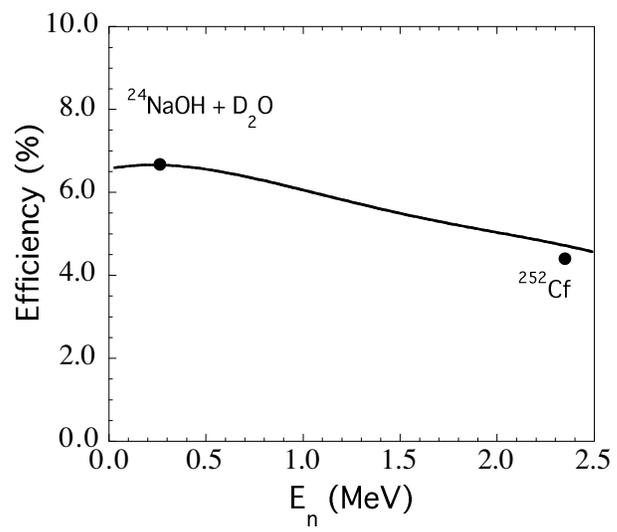}
\end{center}
\caption{Neutron detection efficiency as a function of neutron energy.}
\label{Eff}
\end{figure}

Figure~\ref{Eff} shows the detection efficiency of the neutron detector 
as a function of neutron energy.  
The efficiency was calibrated with 
a 265 keV-neutron source of $^{24}$NaOH + D$_{2}$O and 
a $^{252}$Cf standard source.  
The energy dependence of the efficiency 
was calculated with a 
Monte Carlo code MCNP \cite{Bri97}.  The efficiency was 6.62 \% at 382 keV and 6.65 \% at 156 keV.  
Thus, the difference in detection efficiency between the two reaction channels is negligible.  
Cross sections presented in this paper are based on the detection efficiency for the n + $^{8}$Be reaction channel.

\begin{figure}[t]
\begin{center}
\epsfig{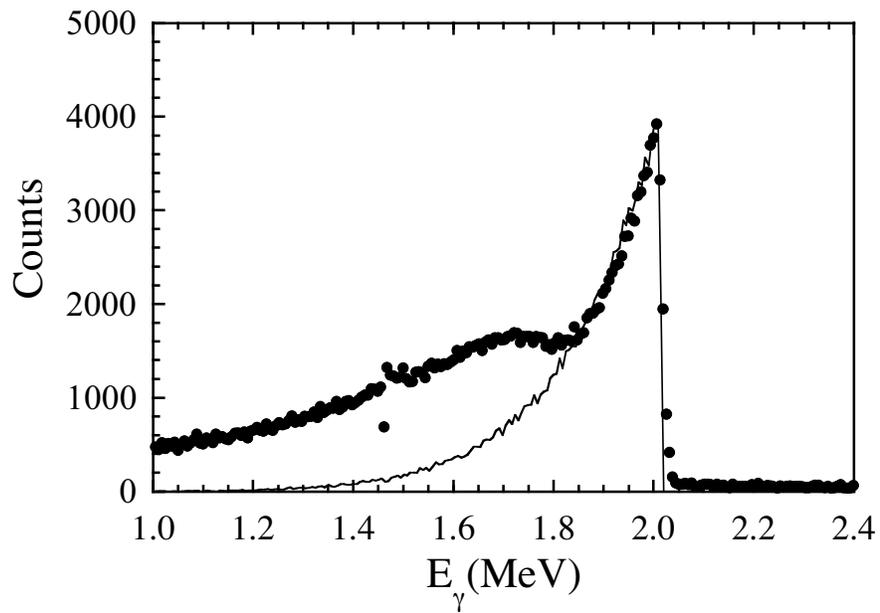}
\end{center}
\caption{Energy distribution of the laser-electron photon beam 
measured with a high-purity Ge detector.  
The energy distribution, 
which best reproduces the response of the Ge detector, 
is shown by the solid curve. }
\label{Ge}
\end{figure}

Figure~\ref{Ge} shows an energy spectrum of the LEPB measured with a 
high-purity Ge detector.  The LEPB has a characteristic low-energy 
tail that is determined by the electron beam emittance, the laser 
optics and the collimator size.  When the 
low-energy tail crosses the neutron threshold of $^{8}$Be + n (1.665 MeV), 
not the total but a partial flux of the LEPB is responsible for 
photodisintegration of $^{9}$Be.  This occurs for LEPBs whose peak
energies are very close to the neutron threshold energy.  Previously photoneutron 
cross sections were presented at energies above 1.78 MeV without flux 
corrections.  As a result, cross sections in the peak region of the 
1/2$^{+}$ state were missing in \cite{Uts01}.  

The flux correction was made for LEPBs 
by means of a Monte Carlo simulation.  Energy 
spectra of the LEPBs taken with the Ge detector were reproduced with 
a Monte Carlo code EGS4 \cite{EGS} by simulating electromagnetic 
interactions between a LEPB 
and the Ge detector.  
The energy 
distribution of the LEPB which best reproduced the response of the Ge 
detector is shown by the solid line in Fig.~\ref{Ge}.  

\begin{figure}[t]
\begin{center}
\epsfig{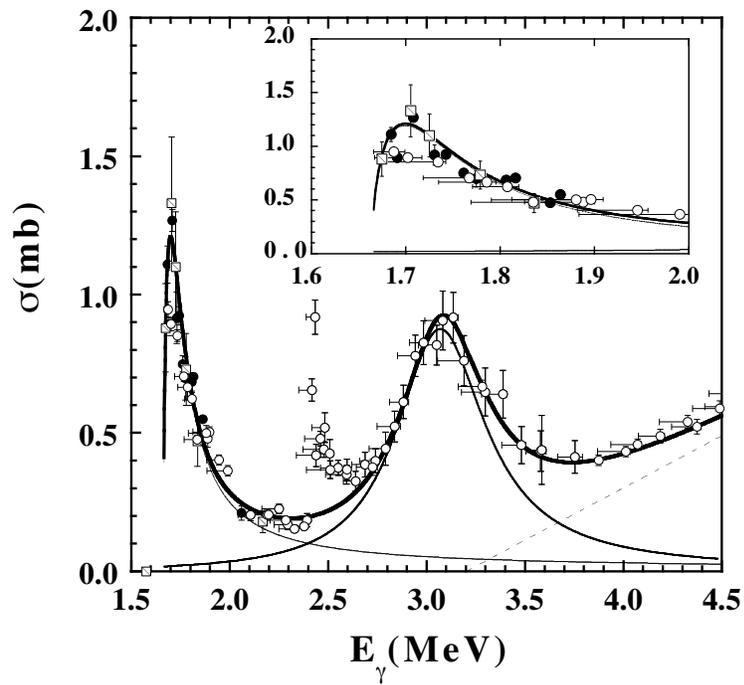}
\end{center}
\caption{Photoneutron cross sections for $^{9}$Be.  A low-energy 
portion is magnified in an inset.  The data below 2 MeV are corrected 
for the $\gamma$-ray flux responsible for the photodisintegration.  
The corrected data are shown by the solid circles.  The data at the 
three lowest energies (1.69, 1.71, 1.73 MeV) near the neutron 
threshold are consistent with the data of a radioactive isotope measurement 
(slashed squares) \cite{Fuj82} as seen in the inset.  The best 
least-squares fit is shown by the solid lines (thick solid line for 
sum, thin solid lines for breakdown).}
\label{Sigma}
\end{figure}

Figure~\ref{Sigma} shows photoneutron cross sections obtained in the 
present experiment.  
The flux correction was made 
for the data in the energy range from the neutron threshold to 2 
MeV.  These data are plotted at the average energies by the solid 
circles in the figure.  Note that the three data points at 1.69, 
1.71, and 1.73 MeV which covered the peak region of the 1/2$^{+}$ 
state are consistent with the data of Fujishiro {\it et al.}  
It was essential to have 
these data points in a new analysis because the statistical uncertainty 
of the Fujishiro data is quite large.  

Including all the data with the flux correction, a new least-squares 
fit was performed to extract the Breit-Wigner parameters for resonance 
states in $^{9}$Be. The same fitting procedure as in \cite{Uts01} was 
followed using a straight-line cross section ($\sigma$ = 0.38 
E$_{\gamma}$ - 1.21 [mb], E$_{\gamma}$ in MeV) for higher-lying 
states in the energy region up to 6.1 MeV.   The E1 parametrization was used for the 
positive-parity states (1/2$^{+}$ and 5/2$^{+}$), while the M1 parametrization for
the negative-parity state (5/2$^{-}$).  
Due to the fact that the energy spread of the LEPBs was much larger than the total width 
($<$ 1 keV) of the 5/2$^{-}$ state, the line shape of the state was not determined experimentally.  
As a result, only $\Gamma_{\gamma}$ was deduced from the integrated cross section.  
The best fit is shown by 
the solid lines.  The best fit parameters are listed in Table 1.  

It is to be pointed out that there are large differences in 
B(E1\,$\downarrow$) and B(M1\,$\downarrow$) for $^{9}$Be states 
between photodisintegration and electron scattering.  
While the present B(E1\,$\downarrow$) for the 1/2$^{+}$ state 
is consistent with the result of the photodisintegration experiment 
with radioactive isotopes (0.106$^{+0.019}_{-0.016}$) \cite{Fuj82}, 
it is larger by a factor of two than the results of electron 
scattering (0.050 $\pm$ 0.020 \cite{Cle68} and 0.054 $\pm$ 0.004 \cite{Kue87}).
Similarly, the present B(E1\,$\downarrow$) for the 5/2$^{+}$ state 
is about 4 times larger than that (0.010 $\pm$ 0.008) 
from (e, e$'$) \cite{Cle68}, whereas the B(M1\,$\downarrow$) 
for the 5/2$^{-}$ state is about half that (0.536 $\pm$ 0.060) 
from (e, e$'$) \cite{Cle68}.  

Regarding the difference for 1/2$^{+}$, Barker previously pointed out a potential problem of 
background subtraction in electron scattering.  As can be seen in the line-shape (Fig.~\ref{Sigma}), 
the 1/2$^{+}$ and 5/2$^{+}$ states overlap each other with substantial tails underneath the narrow 
5/2$^{-}$ state.  The electron scattering strongly enhanced the excitation of the M1 state 
(5/2$^{-}$) adjacent to the E1 states (1/2$^{+}$ and 5/2$^{+}$) \cite{Cle68,Kue87}.  
The authors of Ref. \cite{Cle68} 
realized a background subtraction problem in the 
discussion of the 1/2$^{+}$ state (Section 5.3 of Ref. \cite{Cle68}), 
with the statement 
"Because of the large background and the 
limited statistical accuracy of this small peak, the contribution of the long tail to be 
expected from the ($\gamma$,n) cross section is not included in the result.  Therefore, it is 
reasonable to make a comparison with the ($\gamma$,n) cross section integrated up to 2 MeV 
only."  Indeed, the B(E1\,$\downarrow$) resulting from the present ($\gamma$,n) cross section integrated up 
to 2 MeV is in good agreement with that given in the electron scattering.  Similarly, the 
B(E1\,$\downarrow$) for the weakly populated 5/2$^{+}$ state can be underestimated in the electron 
scattering for the background problem.  This explains, however, the difference only 
qualitatively which is larger for the 5/2$^{+}$ state
(4 times) than for the 1/2$^{+}$ state (2 times).  

Regarding 
the differences for the 5/2$^{-}$ state and the 5/2$^{+}$ state, 
a broad M1 state (1/2$^{-}$), 
which supposedly exists over the 2.5 -- 3.5 MeV region \cite{Chr66}, complicates the situation.   
The present least-squares fit 
did not single out the 1/2$^{-}$ state which was reported to be a 
broad state at 2.78 MeV with $\Gamma \sim$ 1 MeV \cite{Ajz88}.  
If one treats the 1/2$^{-}$ state exactly in the same manner as in the NACRE
compilation (E$_{x}$ = 2.78 $\pm$ 0.12 MeV, $\Gamma$ = 1080 $\pm$ 110 keV,
$\Gamma_{\gamma}$ = 1 W.u. (0.45 eV)), the peak cross section at 2.78 MeV
is found to be 0.066 mb.  This cross section is one order of magnitude smaller 
than the experimental cross section in the corresponding energy region.  
It is interesting to note that the present $\Gamma_{\gamma}$ for 5/2$^{+}$ 
(1.24 $\pm$ 0.02 eV) is accidentally close to the sum 
of the $\Gamma_{\gamma}$'s employed in the NACRE compilation for 
5/2$^{+}$ (0.90 $\pm$ 0.45 eV) and 1/2$^{-}$ (0.45 eV) 
to within the associated uncertainty.  
A more quantitative understanding of this energy region is necessary.  

\subsection{Photoneutron and neutron capture cross sections}\label{sec:sigmas}
We here present analytic formulae to express the cross sections 
with the resonance parameters 
which are determined by the least-squares fit described 
in the previous section.  

The photoneutron cross sections for each resonance state of $^{9}$Be 
are given as,
\begin{equation}
\sigma_{\gamma,n}(1/2^{+}) = 6.654 \times 10^{-2} \bar{E_{\gamma}}
\frac{[181.9(\bar{E_{\gamma}}-1665.3)]^{1/2}}
{[(\bar{E_{\gamma}}-1735.1)^{2}+181.9 (\bar{E_{\gamma}}-1665.3)]}[{\rm mb}],
\end{equation}
\begin{equation}
\sigma_{\gamma,n}(5/2^{+}) = 7.805 \times 10^{-2} \bar{E_{\gamma}}
\frac{274.72}{(\bar{E_{\gamma}}-3077.0)^{2}+75471.}[{\rm mb}],
\label{eq:plus52}
\end{equation}
\begin{equation}
\sigma_{\gamma,n}(5/2^{-}) = 3.4616 \times 10^{7} \bar{E_{\gamma}}^{-2}
\frac{1}{(\bar{E_{\gamma}}-2429.)^{2}+0.14823}[{\rm mb}].
\label{eq:minus52}
\end{equation}
Here, the energy of photon, $\bar{E_{\gamma}}$, is given in unit
of keV.
In the formulae for the broad positive-parity states 
(1/2$^{+}$, 5/2$^{+}$), the energy dependence of 
the $\gamma$-decay width ($\Gamma_{\gamma}$) is explicitly 
taken into account;
\begin{equation}
{\Gamma}_{\gamma} = {16 \pi  \over \mit 9} \alpha  \mit  
{\left({\hbar c}\right)}^{-2} {E}_{\gamma}^{3} \, {\rm B(E1)}, 
\label{eq:gammagamma}
\end{equation}
where $\alpha$ is the fine-structure constant.  
In contrast, Eq. (\ref{eq:minus52}) results from the energy-independent 
${\Gamma}_{\gamma}$ for the sharp 5/2$^{-}$ state.  

The total cross section of photodisintegration of $^{9}$Be 
is the sum of three contributions given by,
\begin{equation}
\sigma_{\gamma,n} = \sigma_{\gamma,n}(1/2^{+})
+ \sigma_{\gamma,n}(5/2^{+}) + f_{n^{8}Be} \, \sigma_{\gamma,n}(5/2^{-}),
\label{eq:sigmagn}
\end{equation}
where the prefactor $f_{n^{8}Be}$ is the branching ratio of 
the $5/2^{-}$ state to the n + $^{8}$Be decay channel.  
We take a 7 \% decay ($f_{n^{8}Be}=0.07$) based on the experimental 
studies \cite{Chr66, Che70} 
to derive the reaction rate.  
The reaction rate does not change largely even if we assume a 
100 \% decay ($f_{n^{8}Be}=1$) to the n + $^{8}$Be decay channel.
This point will be discussed later in section \ref{sec:result}.  

The photoneutron cross section can be converted into 
the neutron capture cross section of $^{8}$Be
based on the detailed-balance theorem.
The neutron capture cross section, $\sigma_{n {\rm Be}}$,
is related with the photoneutron cross section,
$\sigma_{\gamma,n}$, by the formula,
\begin{equation}
\sigma_{n {\rm Be}}(E_{n}) =
\frac{2 (2 j_{a}+ 1)}{(2 j_{b}+ 1)(2 j_{c}+ 1)}
\frac{k_{\gamma}^{2}}{k_{n}^{2}} \sigma_{\gamma,n}(E_{\gamma}),
\end{equation}
where a, b and c denote $^{9}$Be, $^{8}$Be and neutron,
respectively.  
The energy, $E_{n}$, is the neutron energy
with respect to the threshold of n+$^{8}$Be,
defined as,
\begin{equation}
E_{n} = E_{\gamma} - 1.6653,
\label{eq:En}
\end{equation}
in MeV.
The wave numbers 
$k_{n}$ and $k_{\gamma}$ are given, respectively, by,
\begin{equation}
k_{n}^{2} = \frac{2 \mu E_{n}}{\hbar^{2}},
\end{equation}
\begin{equation}
k_{\gamma} = \frac{E_{\gamma}}{\hbar c},
\end{equation}
where $\mu$ is the reduced mass of the n+$^{8}$Be system.
Thus, 
the neutron capture cross section is given by,
\begin{equation}
\sigma_{n {\rm Be}}(E_{n}) = 2.415 \times 10^{-3}
\frac{(E_{n}+1.6653)^{2}}{E_{n}} \sigma_{\gamma,n}(E_{\gamma}),
\label{eq:sigmang}
\end{equation}
from the measured photoneutron cross section,
Eq. (\ref{eq:sigmagn}).
Note that $\bar{E_{\gamma}}$ ($=10^{3}E_{\gamma}$) 
in Eq. (\ref{eq:sigmagn}) is
calculated as $E_{\gamma} = E_{n} + 1.6653$,
through Eq. (\ref{eq:En}).

\section{Theoretical analysis}\label{sec:theory}

\subsection{$\alpha \alpha n$ reaction rate}
We derive the thermal reaction rate of
$\alpha(\alpha n,\gamma)$$^{9}$Be
in two steps through a metastable $^{8}$Be.
We apply the formulation, which was used to derive
the triple alpha reaction by Nomoto {\it et al.} \cite{Nom85},
to the $^{9}$Be formation.
The same formulation is also employed in the NACRE
compilation \cite{Ang99}.

The reaction rate $N_{A}^{2} < \alpha \alpha n >$
(i.e. $N_{A}^{2} < \sigma v >^{\alpha \alpha n}$)
is calculated by,
\begin{equation}
N_{A}^{2} < \sigma v >^{\alpha \alpha n} = N_{A}
\left( \frac{8 \pi }{\mu_{\alpha \alpha}^{2}} \right)
\left( \frac{\mu_{\alpha \alpha}}{2 \pi k_{B} T} \right)^{3/2}
\int_{0}^{\infty} \frac{\hbar \sigma_{\alpha \alpha}(E)}
{\Gamma_{\alpha}(^{8}{\rm Be},E)} \exp (-E/k_{B}T)
N_{A} < \sigma v>^{n ^{8}{\rm Be}} E dE,
\label{eq:aan}
\end{equation}
where $\mu_{\alpha \alpha}$ is the reduced mass of
the $\alpha+\alpha$ system.
$E$ is the energy of $\alpha+\alpha$ with respect to
the threshold (the sum of the rest masses of
$\alpha + \alpha$).
$\sigma_{\alpha \alpha}(E)$ is the cross section
for $^{8}$Be formation at an energy $E$, which
is, in general, different from the ground-state
resonance energy of $^{8}$Be, $E_{^{8}{\rm Be}}$.
$\Gamma_{\alpha}(^{8}{\rm Be},E)$ is the $\alpha$-decay
width of $^{8}$Be at an energy E.
We describe the $^{8}$Be formation and its $\alpha$
decay in the next subsection.

The neutron capture rate of $^{8}$Be is calculated by,
\begin{equation}
N_{A} < \sigma v>^{n ^{8}{\rm Be}} = N_{A}
\left( \frac{8 \pi }{\mu_{n ^{8}{\rm Be}}^{2}} \right)
\left( \frac{\mu_{n ^{8}{\rm Be}}}{2 \pi k_{B} T} \right)^{3/2}
\int_{0}^{\infty} \sigma_{n ^{8}{\rm Be}}(E';E)
\exp (-E'/k_{B}T) E' dE',
\label{eq:n8be}
\end{equation}
where $\mu_{n ^{8}{\rm Be}}$ is the reduced mass of
the n + $^{8}$Be system.
$E'$ is the energy of n + $^{8}{\rm Be}$ with respect to
the threshold (the sum of the rest masses of a neutron
and a $^{8}$Be).
Note that this threshold varies with the formation
energy $E$ of $^{8}$Be.
$\sigma_{n ^{8}{\rm Be}}(E';E)$ is the cross section
for neutron capture of $^{8}$Be at the formation energy $E$.

We calculate this cross section using the result of
Eq. (\ref{eq:sigmang}) in section \ref{sec:sigmas}.
Since the neutron energy in Eq. (\ref{eq:sigmang})
is measured from the threshold for n + $^{8}{\rm Be}$, 
we convert $E'$ to $E_{n}$ 
with the ground-state resonance energy $E_{^{8}{\rm Be}}$
by,

\begin{equation}
E_{n} = E' + E - E_{^{8}{\rm Be}},
\end{equation}
to calculate the neutron capture cross section.

\subsection{$^{8}$Be formation}

We calculate the $^{8}$Be formation cross section 
following the prescription of Nomoto {\it et al.}
For details, see Section 2.2.2 of Ref. \cite{Nom85}.
The cross section is given by,
\begin{equation}
\sigma_{\alpha \alpha}(E) = \frac{S(E)}{E}
\exp [ - (E_{G}/E)^{1/2}],
\label{eq:sigmaaa}
\end{equation}
where $E_{G}$ is the Gamow energy given by,
\begin{equation}
E_{G}^{1/2} = 0.98948 Z_{1} Z_{2} A^{1/2} \, {\rm MeV}^{1/2}.
\end{equation}
Here $Z_{1}$, $Z_{2}$ are the proton numbers of
target and projectile nuclei.
$A$ is the reduced mass number,
\begin{equation}
A = \frac{A_{1}A_{2}}{A_{1}+A_{2}},
\end{equation}
calculated from the mass numbers 
of target and projectile nuclei $A_{1}$, $A_{2}$.

The S-factor $S(E)$ is calculated by,
\begin{equation}
S(E) = \frac{0.6566 \, \omega_{r}}{A}
\frac{\Gamma_{1}(E_{r}) \Gamma_{2}(E)}
{(E-E_{r})^{2} + \Gamma(E)^{2}/4} \,
\exp [ (E_{G}/E_{r})^{1/2} + \alpha_{l} E_{r} - \alpha_{l} E)]
\, {\rm MeV \, barn},
\label{eq:sfac}
\end{equation}
where $E_{r}$ is the resonance energy and
$\omega_{r}=[(2J_{r}+1)/(2J_{1}+1)(2J_{2}+1)](1+\delta_{12})$.
The energy dependent total decay width $\Gamma(E)$ is
calculated by,
\begin{equation}
\Gamma(E) = \Gamma(E_{r}) \,
\exp [ - (E_{G}/E)^{1/2} - \alpha_{l}E
+ (E_{G}/E_{r})^{1/2} + \alpha_{l}E_{r}].
\end{equation}
Note that $\Gamma_{2}(E)$ in Eq. (\ref{eq:sfac}) 
cancels out with 
$\Gamma_{\alpha}(^{8}{\rm Be},E)$ in Eq. (\ref{eq:aan}).  
Taking $l=0$, 
the factor $\alpha_{l}$ is defined
by \cite{Bur57},
\begin{equation}
\alpha_{l=0} = \frac{1}{3 E_{R}}
\left\{ \frac{K_{3}(x)}{K_{1}(x)} - 1 \right\},
\end{equation}
where 
$K_{l}(x)$ is the modified Bessel function of order $l$,
$E_{R}$ is given by,
\begin{equation}
E_{R} = \frac{20.9}{A R^{2}} \, {\rm MeV}
\end{equation}
with $R=1.44(A_{1}^{1/3}+A_{2}^{1/3})$ fm and 
$x$ is given by
\begin{equation}
x = 0.525 (A Z_{1} Z_{2} R)^{1/2}.
\end{equation}

\section{Numerical results}\label{sec:result}
\subsection{Calculated reaction rate}
We numerically integrate Eqs. (\ref{eq:aan}) and 
(\ref{eq:n8be}) with the cross sections 
Eqs. (\ref{eq:sigmang}) and (\ref{eq:sigmaaa})
to derive the reaction rate 
$N_{A}^{2} < \alpha \alpha n >$.
We provide the calculated data of the 
reaction rate $N_{A}^{2} < \alpha \alpha n >$ 
in Table \ref{table:table2}.  
The grid points of temperature is the same as those 
of the NACRE compilation for comparison.  

\begin{figure}[t]
\begin{center}
\epsfig{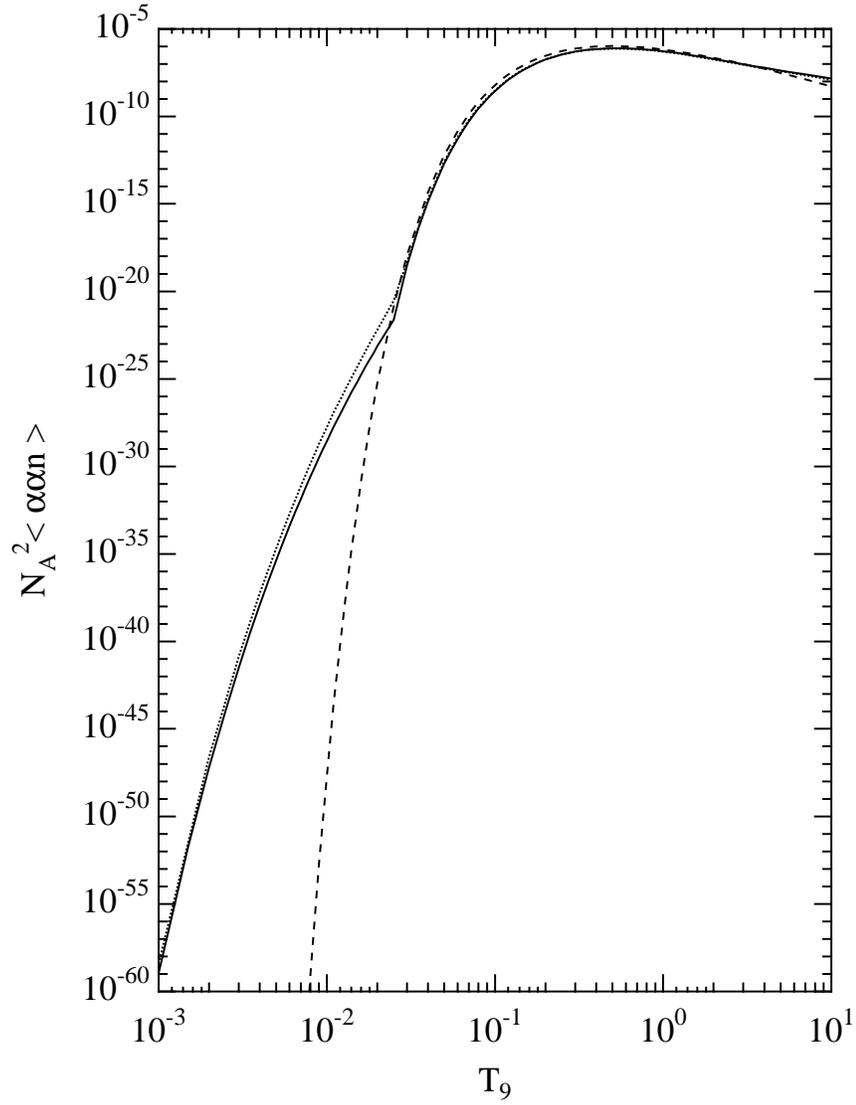}
\end{center}
\caption{The reaction rate $N_{A}^{2} < \alpha \alpha n >$ in the present 
study is shown by the solid curve as a function of temperature.
The reaction rates by CF88 \cite{Cau88} 
and NACRE \cite{Ang99} are shown by
dashed and dotted curves, respectively.}
\label{aan1}
\end{figure}

\begin{figure}[t]
\begin{center}
\epsfig{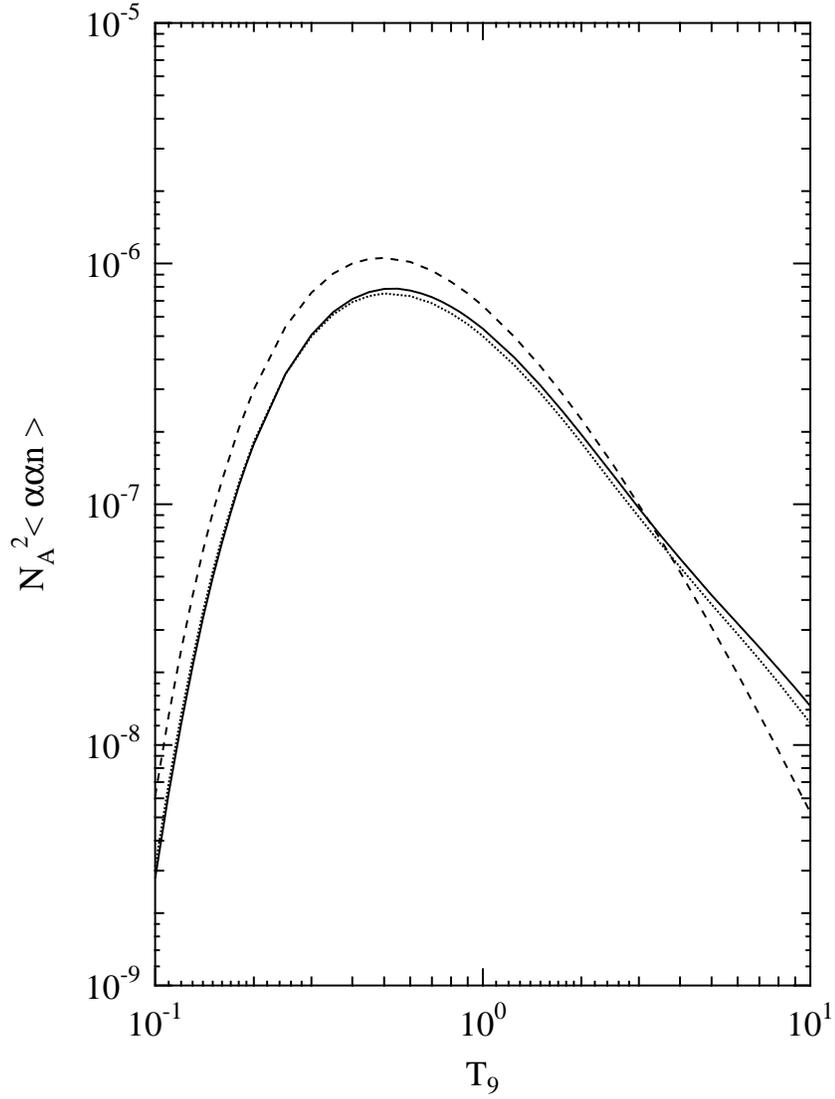}
\end{center}
\caption{The zoom up of Fig. \ref{aan1} for a high temperature range.
The reaction rates $N_{A}^{2} < \alpha \alpha n >$ are shown
as functions
of temperature between $T_{9}=10^{-1}$ and $T_{9}$=10 
with the same notation as in Fig. \ref{aan1}.}
\label{aan2}
\end{figure}

Figures \ref{aan1} and \ref{aan2} display the numerical result 
of the reaction rate $N_{A}^{2} < \alpha \alpha n >$ 
as a function of temperature ($T_{9}$).
For comparison, 
the reaction rates by Caughlan and Fowler (CF88) \cite{Cau88} 
and Angulo {\it et al.} (NACRE) \cite{Ang99} are shown 
by dashed and dotted curves, respectively.  

The difference from the CF88 is apparent at $T_{9} \leq 0.028$ 
due to lack of the off-resonant contribution in CF88.  
The neglect of the off-resonant contribution 
leads to the underestimate of the reaction rate below $T_{9}=0.01$
by more than 20 orders of magnitude.  
The proper treatment of the off-resonant contribution 
from the metastable $^{8}$Be nucleus 
is, therefore, essential at low temperatures.  
The difference remains substantial at high temperatures 
as seen in Figure \ref{aan2}.  
The CF88 rate is larger than the present rate 
at $0.028 < T_{9} < 3$ and becomes smaller above $T_{9} \sim 3$.
On the other hand, the NACRE rate is significantly 
larger than the present rate 
at low temperatures $T_{9} \leq 0.028$ (Fig. \ref{aan1}), 
whereas it is in good agreement with the present 
rate at high temperatures $T_{9} \geq 0.1$ (Fig. \ref{aan2}).

\begin{figure}[t]
\begin{center}
\epsfig{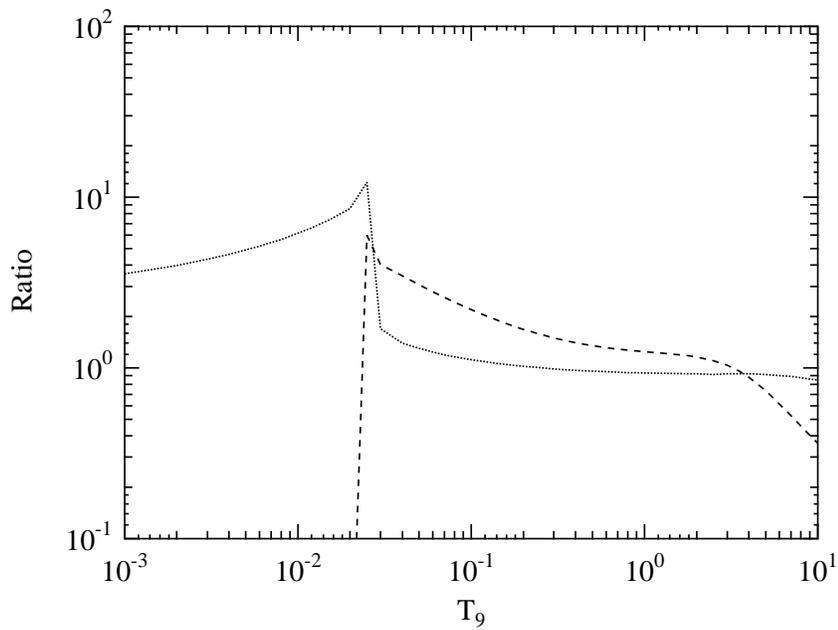}
\end{center}
\caption{The ratios of the reaction rate $N_{A}^{2} < \alpha \alpha n >$ 
of the CF88 and NACRE compilations to the present reaction rate 
are shown by dashed and dotted curves, respectively, 
as functions of temperature.}
\label{ratio}
\end{figure}

The ratios of the CF88 rate and the NACRE rate 
to the present reaction rate are shown 
in Fig. \ref{ratio} by the dashed curve and the dotted curve, 
respectively.  
The difference of the CF88 from the present rate 
amounts to a factor of 2 at $T_{9} \geq 0.1$.  
This is because 
the CF88 is based on the photoneutron cross section of Berman {\it et al.}, 
which shows a larger cross section (1.6 mb) peaked at an energy 
(1.671 MeV) closer to the threshold than the present cross section.  
At $T_{9} \geq 3$, the CF88 rate becomes smaller 
because it neglects other resonances 
(mainly the 5/2$^{-}$ state) at higher energies.  
The NACRE rate is larger by factors 4--12 than the present rate 
in the low temperature regime ($T_{9} < 0.028$).  
It becomes closer to the present rate in the transitional 
regime ($T_{9} \sim 0.028$) from off-resonance to on-resonance 
and consistent with the present rate to within $\pm 20 \%$ above $T_{9} = 0.1$.  

The discrepancy between the NACRE rate and the present rate 
at $T_{9} < 0.028$ is attributed to different behaviors of low-energy tails 
of broad resonances.  
Presently, the cross section for the 5/2$^{+}$ state is formulated 
with the energy-dependent ${\Gamma}_{\gamma}$ as in Eq. (\ref{eq:plus52}), 
while the NACRE compilation adopted energy-independent widths 
for the same state (${\Gamma}_{\gamma}$ = 0.90 $\pm$ 0.45 eV) and, 
in addition, for the 1/2$^{-}$ state 
(${\Gamma}_{\gamma}$ = 1 W.u. = 0.45 eV).
As a result, the low-energy tail decreases faster with energy 
in the present cross section than that in the NACRE compilation.  

The 1/2$^{+}$ resonance immediately above the neutron threshold
plays an essential role in the reaction rate.  
The present photoneutron cross section for the 1/2$^{+}$ resonance 
agrees well with that of Fujishiro {\it et al.} \cite{Fuj82} 
which was employed in the NACRE compilation.  
This explains a similarity between the present rate and the 
NACRE rate at $T_{9} \geq 0.1$.  

The 5/2$^{-}$ resonance 
contributes to the reaction rate only 
at high temperatures.  
In the current study, 
we have taken the 7 $\%$ decay of the $5/2^{-}$ state 
to the n + $^{8}$Be channel.  
To see the sensitivity to the branching ratio, 
we calculated the reaction 
rate assuming the 100 $\%$ decay ($f_{n^{8}Be}=1$ 
in Eq. (\ref{eq:sigmagn})).  
It was found that the 100 $\%$ branching does not change 
the reaction rate below $T_{9}$=1 
and increases the rate only by 5 $\%$ around $T_{9}$=10.  

\subsection{Analytical expressions}
We provide the analytical
formula which reproduces the numerical result of the reaction rate 
$N_{A}^{2} < \alpha \alpha n >$ 
given in Table \ref{table:table2}.
We fit the numerical data by using 
the same formulae as those adopted 
in Table 3 of the NACRE compilation \cite{Ang99}.  
The reaction rate of the $^{8}$Be 
formation, $N_{A} < \alpha \alpha >_{formula}$, 
is expressed by 
\begin{eqnarray}
N_{A}< \alpha \alpha >_{formula} &=& 2.43 \times 10^{9} \, T_{9}^{-2/3} 
\exp ( -13.490 \, T_{9}^{-1/3} - (T_{9}/0.15)^{2}) 
\times (1 + 74.5 \, T_{9}) \nonumber \\
&& + 6.09 \times 10^{5} \, T_{9}^{-3/2} \exp (-1.054/T_{9}).
\end{eqnarray}
The fitting formula of the 
reaction rate $N_{A}^{2} < \alpha \alpha n >$ for $T_{9} \leq 0.03$ 
is parameterized as 
\begin{equation}
N_{A}^{2}< \alpha \alpha n >_{fitting} = N_{A}< \alpha \alpha >_{formula} \times 
a_{1} (1 + a_{2} T_{9}     + a_{3} T_{9}^{2} 
         + a_{4} T_{9}^{3} + a_{5} T_{9}^{4}
         + a_{6} T_{9}^{5}),
\end{equation}
and for $T_{9} > 0.03$ as 
\begin{equation}
N_{A}^{2}< \alpha \alpha n >_{fitting} = N_{A}< \alpha \alpha >_{formula} \times 
a_{1}  (1 + a_{2}  \log_{10} T_{9}    + a_{3} (\log_{10} T_{9})^{2} 
          + a_{4} (\log_{10} T_{9})^{3}),
\end{equation}
where $a_{i}$ are coefficients to be determined.  
We perform a least-squares fit 
to the numerical data of Table \ref{table:table2}.  
The best fit formula for $T_{9} \leq 0.03$ is 
\begin{eqnarray}
N_{A}^{2}< \alpha \alpha n >_{fitting}&=&N_{A}< \alpha \alpha >_{formula} \times 
1.376 \times 10^{-12} \nonumber \\
&& \times (1 -58.80 \, T_{9} -1.794 \times 10^{4} \, T_{9}^{2} 
+ 2.969 \times 10^{6} \, T_{9}^{3} \nonumber \\
&& -1.535 \times 10^{8} \, T_{9}^{4} + 2.610 \times 10^{9} \, T_{9}^{5}),
\end{eqnarray}
and for $T_{9} > 0.03$ is
\begin{eqnarray}
N_{A}^{2}< \alpha \alpha n >_{fitting}&=&N_{A}< \alpha \alpha >_{formula} \times 
2.630 \times 10^{-12} \times (1 -1.359 \log_{10} T_{9} \nonumber \\
&& +1.995 \times 10^{-1} (\log_{10} T_{9})^{2} + 5.251 \times 10^{-1} (\log_{10} T_{9})^{3}).
\end{eqnarray}
The above formulae reproduce the numerical data 
of the reaction rate in Table \ref{table:table2} 
within $\sim$ 20 $\%$ deviation.  

\section{Summary and discussions}\label{sec:summary}
We have studied the astrophysical reaction rate for 
$\alpha(\alpha n,\gamma)^{9}$Be, which is important 
in nucleosynthesis under alpha- and neutron-rich 
environments.  
The formation of $^{9}$Be by this three body reaction 
plays an essential role to bridge the mass gap at A=8 and to 
create intermediate-to-heavy mass nuclei starting 
from light elements.  
This reaction rate influences largely 
the alpha-rich freeze-out followed by the r-process 
nucleosynthesis in neutrino-driven winds and possibly 
other nucleosynthesis in various astrophysical sites.  

We adopted the latest experimental data 
on the photodisintegration of $^{9}$Be taken 
with laser-electron photon beams.  
This experiment covered all resonance states 
of astrophysical importance in $^{9}$Be.  
We added data points near the neutron 
threshold with the flux correction for incident photons
and updated the least-squares analysis of extracting 
resonance parameters.  

We numerically calculated the thermal 
average of cross sections for the formation 
of $^{9}$Be via the metastable $^{8}$Be, 
i.e., $\alpha + \alpha$ $\rightleftharpoons$ $^{8}$Be(n,$\gamma$)$^{9}$Be.  
We took into account both off-resonant and 
on-resonant contributions from 
the ground state in $^{8}$Be.  
The off-resonant contribution, 
which was not included in the previous 
brief report \cite{Uts01} on the photodisintegration experiment, 
becomes important at low temperatures below $T_{9}=0.028$.  

We have provided the reaction rate in the wide 
temperature range 
from T$_{9}$=10$^{-3}$ to T$_{9}$=10$^{1}$ both in 
a tabular form and in analytic formulae.  
The calculated reaction rate is compared 
with the reaction rates of the CF88 and the NACRE compilations.  
The CF88 rate, which is invalid at $T_{9} \leq 0.028$
due to lack of the off-resonant contribution, 
differs from the present reaction rate by a factor of 2 
at high temperatures $T_{9} \geq 0.1$.  
The NACRE rate is 4--12 times larger than 
the present reaction rate 
at low temperature ($T_{9} < 0.028$) 
and becomes closer to the present rate 
in the transitional temperature region 
from off-resonant to on-resonant processes.  
The NACRE rate is consistent with the present rate 
at $T_{9} \geq 0.1$ to within $\pm 20 \%$.  
Because of the systematic treatment of photoneutron 
cross sections obtained from the dedicated experiment, 
we would like to recommend to use the new reaction 
rate for possible astrophysical problems.  

It is worthwhile mentioning possible impacts of the 
new reaction rate on modeling nucleosynthesis in various 
astrophysical sites.  
The production of seed elements in the neutrino-driven 
wind and the alpha-rich freeze-out scenarios can be 
enhanced 
due to the slight increase of the reaction rate 
at $T_{9} > 3.5$.  
This may lead to a slightly smaller neutron-to-seed 
ratio at the onset of r-process, and hence make 
an appreciable influence on the entropy condition and 
expansion dynamics of the neutrino-driven wind \cite{Ots00}.  
Kajino {\it et al.} have recently studied the sensitivity 
of the r-process yields 
to the $\alpha(\alpha n,\gamma)^{9}$Be reaction rate, 
and found that a factor of two larger rate would 
change the final r-process abundances 
drastically in rapid expansion models of 
the neutrino-driven wind \cite{Kaj02}.
These possible implications should be studied in detail 
by using the present new reaction rate.  
It is to be clarified quantitatively 
if the $\alpha(\alpha n,\gamma)^{9}$Be reaction path 
still predominates the seed production 
and to what extent 
the final r-process yields may change by including 
the present reaction 
rate and also newly-identified competing reactions 
$\alpha(t,\gamma)^{7}$Li$(n,\gamma)^{8}$Li$(\alpha,n)^{11}$B 
\cite{Ter01,Ter01a} in the nuclear reaction network.

It is also postulated theoretically that the production 
of $^{9}$Be in the He layer of AGB stars may change 
the neutron-rich environment required for the 
s-process nucleosynthesis.  Once proton mixing 
into C-rich He layer could occur either in the 
interpulse phase \cite{Gal98, Bus99} 
or in the thermal pulse phase \cite{Iwa01} 
of the AGB stars, the reignition of H-burning 
proton capture on abundant $^{12}$C would form $^{13}$C.  
Temperature might exceed $T_{9}=0.09$ so that 
$^{13}$C$(\alpha,n)^{16}$O provides plenty of free neutrons.  
Since this neutron production could occur in the He layer, 
the $\alpha(\alpha n,\gamma)^{9}$Be reaction 
may follow 
and play a role of neutron-poison which regulates the 
s-process nucleosynthesis.  
As the burning time scale of $\alpha \alpha n$ depends 
strongly on the local neutron density and temperature, 
it is highly desirable to calculate precise 
profile of the neutron exposure and the temperature distribution 
in order to identify above theoretical speculation.  
Note that, should $\alpha(\alpha n,\gamma)^{9}$Be reaction 
occur in such an environment of AGB stars, it is hard to observe $^{9}$Be 
because $^{9}$Be is easily destroyed by the $(p,\alpha)$ reaction 
in the next thermal pulse.  

Another neutron-rich site in cosmological 
nucleosynthesis is baryon inhomogeneous Big-Bang Universe 
\cite{Ori97}.  
In segregated fluctuations of neutron and proton number-density 
distributions it is identified that the 
$^{7}$Li$(n,\gamma)^{8}$Li$(\alpha,n)^{11}$B and 
$^{7}$Li$(t,n)^{9}$Be reactions are 
the most viable nuclear flows to produce 
intermediate-to-heavy mass nuclei in the Big-Bang nucleosynthesis 
\cite{App88, Kaj90, Boy01}.  
It should be studied theoretically 
if the $\alpha(\alpha n,\gamma)^{9}$Be reaction can be an alternative 
flow path or a competing reaction with those identified 
previously.  

The $\alpha(\alpha n,\gamma)^{9}$Be stellar reaction rate was 
evaluated in a more consistent manner, 
where the systematic experimental 
data of photodisintegration of $^{9}$Be was used, 
than in the preceding compilations \cite{Cau88,Ang99}.  
As of the present evaluation, however, there still remains some uncertainty 
regarding the contribution from the $^{5}$He + $\alpha$ channel 
due to a lack of experimental information on low-lying states in $^{9}$Be.  
A further experimental investigation with emphasis 
on the $^{5}$He + $\alpha$ configuration is advisable.

\section*{Acknowledgment}
The authors are grateful for continuous collaborations
and fruitful discussions with M. Terasawa, K. Otsuki
and S. Wanajo.
K. S. and T. K. thank for the hospitality of Konan
University, where a part of this work has been done.
This work was in part supported by the Japan Private 
School Promotion Foundation, the HIRAO TARO Foundation 
of Konan University Association for Academic Research, 
and 
the Japan Society for Promotion of Science 
under the Grants-in-Aid Program for Scientific Research 
(13740165, 13640303).  


\newpage

\begin{table}[t]
\caption{Resonance parameters deduced in the present experiment.}
\label{table:table1}
\begin{center}
\[\begin{tabular}{cccccc} \hline
$I^{\pi}$ & $X \lambda$ & $E_{R}$ & $B(X \lambda \downarrow)$ & $\Gamma_{\gamma}$ & $\Gamma_{n}$\\
          &    & [MeV]             & E1:[$e^{2}fm^{2}$]       &           [eV]    &     [keV]   \\
          &    &                   & M1:[$(e\hbar/2Mc)^{2}$]  &                   &             \\ \hline
$1/2^{+}$ & E1 & $1.735 \pm 0.003$ & $0.104  \pm 0.002$       & $0.568 \pm 0.011$ & $225 \pm 12$\\ \hline
$5/2^{-}$ & M1 & $2.43           $ & $0.295  \pm 0.072$       & $0.049 \pm 0.012$ &             \\ \hline
$5/2^{+}$ & E1 & $3.077 \pm 0.009$ & $0.0406 \pm 0.0007$      & $1.24  \pm 0.02$  & $549 \pm 12$\\ \hline
\end{tabular} \]
\end{center}
\end{table}

\begin{table}[t]
\caption{The thermal reaction rate $N_{A}^{2}< \alpha \alpha n >$
at representative temperatures.}
\label{table:table2}
\begin{center}
\begin{footnotesize}
\begin{tabular}{cc|cc} \hline
$T_{9}$ & rate & $T_{9}$ & rate \\ \hline
0.001 	&	1.10E-59	&	0.14 	&	3.35E-08	\\
0.002 	&	6.30E-48	&	0.15 	&	4.96E-08	\\
0.003 	&	3.12E-42	&	0.16 	&	6.92E-08	\\
0.004 	&	1.21E-38	&	0.18 	&	1.18E-07	\\
0.005 	&	4.31E-36	&	0.2 	&	1.78E-07	\\
0.006 	&	3.81E-34	&	0.25 	&	3.47E-07	\\
0.007 	&	1.36E-32	&	0.3 	&	5.06E-07	\\
0.008 	&	2.62E-31	&	0.35 	&	6.29E-07	\\
0.009 	&	3.19E-30	&	0.4 	&	7.13E-07	\\
0.01 	&	2.75E-29	&	0.45 	&	7.62E-07	\\
0.011 	&	1.81E-28	&	0.5 	&	7.83E-07	\\
0.012 	&	9.63E-28	&	0.6 	&	7.73E-07	\\
0.013 	&	4.30E-27	&	0.7 	&	7.25E-07	\\
0.014 	&	1.66E-26	&	0.8 	&	6.62E-07	\\
0.015 	&	5.68E-26	&	0.9 	&	5.97E-07	\\
0.016 	&	1.75E-25	&	1 	&	5.36E-07	\\
0.018 	&	1.30E-24	&	1.25 	&	4.06E-07	\\
0.02 	&	7.33E-24	&	1.5 	&	3.12E-07	\\
0.025 	&	2.45E-22	&	1.75 	&	2.44E-07	\\
0.03 	&	2.98E-19	&	2 	&	1.95E-07	\\
0.04 	&	1.36E-15	&	2.5 	&	1.32E-07	\\
0.05 	&	1.98E-13	&	3 	&	9.61E-08	\\
0.06 	&	5.20E-12	&	3.5 	&	7.39E-08	\\
0.07 	&	5.15E-11	&	4 	&	5.94E-08	\\
0.08 	&	2.79E-10	&	5 	&	4.20E-08	\\
0.09 	&	1.01E-09	&	6 	&	3.19E-08	\\
0.1 	&	2.79E-09	&	7 	&	2.54E-08	\\
0.11 	&	6.29E-09	&	8 	&	2.07E-08	\\
0.12 	&	1.22E-08	&	9 	&	1.72E-08	\\
0.13 	&	2.12E-08	&	10 	&	1.45E-08	\\ \hline
\end{tabular}
\end{footnotesize}
\end{center}
\end{table}

\newpage

\end{document}